\documentstyle[preprint,aps]{revtex}

\def\vareps{\varepsilon}
\begin{document}
\draft
%
%
\title{Theory for the Nonlinear Optical Response of Transition Metals:
Polarization Dependence as a Fingerprint of the Electronic Structure
at Surfaces and Interfaces}
\author{W. H\"ubner, K. H. Bennemann, and K. B\"ohmer}
\address{Physics Department, Freie Universit\"at
Berlin, Arnimallee 14, D-14195 Berlin, Germany}
\date{\today}
\maketitle
\begin{abstract}
We show that the nonlinear optical response reflects sensitively the electronic
structure of transition-metal surfaces  and interfaces. $d$ and $s$ electrons
may contribute rather differently to the second harmonic generation (SHG)
signal.
This results from the different sensitivity of $d$ and $s$ electrons to
surface and symmetry changes. Consequently, SHG for noble metals shows a
by far larger dependence on the polarization of the incoming light than
for transition metals like Fe, Co, Ni, in particular at lower frequencies.
The theoretical results are compared with
recent measurements. We conclude that the SHG yield is in addition to the
nonlinear magneto-optical Kerr-effect a sensitive fingerprint of the
electronic structure at surfaces and interfaces.
\end{abstract}
\newpage
\noindent
{\bf I. Introduction}\\
\indent
Recently, the nonlinear optical response of metallic surfaces
has been studied intensively. The theoretical analysis uses typically the
free-electron approximation or the jellium model~\cite{bloem,liebsch}.
Clearly, this must be extended for studying the nonlinear optical response of
transition metals and for analyzing $d$- vs. $s$-electron contributions to
optical second harmonic generation (SHG). Here, we present such a theory by
treating the electrons using a tight-binding like model which has been
successful in analyzing the nonlinear magneto-optical Kerr-effect
(NOLIMOKE)~\cite{hub}. As a result of the different localization of
$s$- and $d$- wavefunctions and associated different sensitivity
with respect to surfaces and interfaces and symmetry changes, we find
that the wavelength and polarization dependence of the SH light on the
incoming light is rather different for noble metals such as Cu, Ag, Au and for
transition metals such as Ni, Co, Fe, and Cr, in particular at lower
frequencies.

Denoting by $I_{\alpha }(\beta -SH)$ the yield of $\beta $-polarized
second harmonic light resulting from $\alpha $-polarized
incoming light, then quite general theoretical arguments
predict that for transition metals involving $d$-electron
transitions
\begin{equation}
I_{p}(p-SH)\stackrel{>}{\sim} I_{s}(p-SH)\;,
\end{equation}
while
\begin{equation}
I_{p}(p-SH)\gg I_{s}(p-SH)\approx 0\;,
\end{equation}
if only $s$-$p$ electronic
transitions are occurring. Hence, for noble metals
$I_{p}(p-SH)\gg I_{s}(p-SH)\approx 0$ for light frequencies
$\omega $ which do not permit $d$-electronic transitions,
e. g.  $\hbar\omega \;<\;(\varepsilon_{F}\;-\;\varepsilon_{d}^{max})$.
Here, $\varepsilon_{F}$ is the Fermi energy and $\varepsilon_{d}^{max}$
refers to the upper edge of the occupied $d$-band. However, for
$\hbar\omega \;\geq\;(\varepsilon_{F}\;-\;\varepsilon_{d}^{max})$
noble metals will behave more and more like transition metals.

This behavior results from the fact that for $s$ (and $p$) electrons
one gets for the dipole matrix elements controlling the nonlinear
optical response $\langle s\mid x\mid s \rangle\;\approx \; 0$
and
\newline
$\langle s\mid z\mid s \rangle\;\neq \; 0$, and furthermore
$\langle d\mid z\mid d \rangle\;\geq \;\langle d\mid x\mid d \rangle$
for transitions involving $d$-electrons. The coordinates $z$ and $x$
refer to directions perpendicular and parallel to the surface, see Fig. 1.
This behavior of the
nonlinear optical response is in agreement with recent
experimental observations~\cite{petro,klaus,klausdiss}.
To understand the effect of the matrix elements on the nonlinear optical
response in more detail, note
that the nonlinear optical response is given by
\begin{equation}
P_{i}\;=\;\chi^{(2)}_{ijl}E_{j}E_{l},
\end{equation}
where $P_{i}$ is the i-th component of the polarizability and where the
susceptibility tensor $\chi^{(2)}_{ijl}$ is determined within
an electronic theory using response theory~\cite{hub}.
Thus, the SHG yield is determined by
\begin{equation}
I_{\alpha }(\beta -SH)\sim \mid\sum_{ijl}\chi_{ijl}^{(2)}\mid^{2}
\end{equation}
neglecting for simplicity at the moment Fresnel and transmission coefficients.
The nonlinear susceptibility is now of the form
\begin{equation}
\chi^{(2)}_{ijl} \;=\; \sum_{k k^{\prime } k^{\prime
\prime }}\langle k\mid i\mid k^{\prime }\rangle\;
\langle k^{\prime }\mid j\mid k^{\prime \prime}\rangle\;\langle
k^{\prime \prime }\mid l\mid k\rangle\;F_{k k^{\prime } k^{\prime
\prime }}(\omega ,2\omega )\;,
\end{equation}
where $\langle k\mid i\mid k^{\prime }\rangle$ is the usual dipole-matrix
element involving electronic transitions between $s$, $p$, and $d$ states
$k$ and $k^{\prime }$.
Note, as a result of inversion symmetry breaking at surfaces and interfaces
$s\longrightarrow (s,p)$, $s\longrightarrow d$, and $d\longrightarrow d$
transitions occur~\cite{note}. The function $F_{k k^{\prime } k^{\prime
\prime }}(\omega )$ collects all other
factors (energy denominators and Fermi functions) and gives essentially the
doubly convoluted density of states~\cite{note1}. Decomposing
$\chi^{(2)}_{ijl}$ with respect to the electronic transitions one gets
\begin{equation}
\chi^{(2)}_{ijl}\;=\;\chi^{(2)s}_{ijl}\;+\;\chi^{(2)d}_{ijl}
\;+\;\chi^{(2)int}_{ijl},
\end{equation}
with $\chi^{(2)int}_{ijl}$ involving  only $s,p\longrightarrow d$ transitions,
while  $\chi^{(2)s}_{ijl}$ involves only $s\longrightarrow s$ transitions
and $\chi^{(2)d}_{ijl}$ involves  only $d\longrightarrow d$ transitions.
It can be shown that due to the dipole matrix elements the various
contributions to the tensor elements $\chi^{(2)}_{ijl}$ are of different
magnitude, see table I. This will simplify the analysis.

Using now for a simplified treatment of the wave functions the atomic basis
set of ref.~\cite{hub}, we calculate the dipole
matrix elements. Results are listed in table I. Thus, we get typically
for transition metals $\chi^{(2)d}_{ijl} \gg\chi^{(2)s}_{ijl}$, and for
noble metals $\chi^{(2)}_{ijl}\approx\chi^{(2)s}_{ijl}$,
if $\hbar \omega \;<\;(\varepsilon_{F}-\varepsilon_{d}^{max})$. As a
consequence, one
estimates for noble metals with
$\hbar \omega \;<\;(\varepsilon_{F}-\varepsilon_{d}^{max})$ that
due to $\langle s\mid x\mid s \rangle\;\approx \; 0$:
\begin{equation}
I_{p}(p-SH)\gg I_{s}(p-SH)\;,\;I_{s}(p-SH)\approx 0\;.
\end{equation}
Furthermore, for transition metals we estimate:
\begin{equation}
I_{\alpha }(\beta -SH)\sim \mid\sum_{ijl}\chi_{ijl}^{(2)d}\mid^{2}\;,
\end{equation}
and thus
\begin{equation}
I_{p}(p-SH)\;\stackrel{>}{\sim}\; 3 I_{s}(p-SH),
\end{equation}
since $\langle d\mid z\mid d \rangle\;\stackrel{>}{\sim}\;\langle d\mid x\mid d
\rangle$,
see table I.

This behavior of the second harmonic intensity $I_{\alpha }(\beta -SH)$ is
in accordance with recent
experiments~\cite{petro,klaus,klausdiss} and shows already clearly
that the SHG yield reflects sensitively the electronic structure
of noble metals and transition metals at surfaces. Of course, this is
of great interest not only for surface studies, but also for the
investigation of thin films and interfaces. Note, for a more-detailed
study of the polarization dependence of $I_{\alpha }(p-SH)$,
with $\alpha $ varying from $p$ to $s$, Fresnel and transmission
coefficients have to be included to determine the SHG yield.\\
In section II we present a simplified theory for the polarization
dependence of the nonlinear optical
response which may serve as a basis also for further extensive studies.
In section III
we present results and compare with recent experiments.
\\
\\
{\bf II. Theory}\\
That the SHG yield is a sensitive fingerprint of the electronic
structure at surfaces and interfaces becomes obvious from the
polarization dependence of the SHG light intensity
$I_{\alpha }(\beta -SH)$. Here, $\alpha $ refers to the polarization
of the incoming light.
In order to determine more accurately and in detail this intensity
$I_{\alpha }(\beta -SH )$ of the $\beta $-polarized SHG light at frequency
2$\omega $, note that
the reflected light at frequency $2\omega $ is given as shown by
B\"ohmer {\em et al.}~\cite{klaus,klausdiss} for C$_{4v}$ symmetry
without magnetization by
\newpage
\begin{eqnarray}
&&E^{(2\omega)}(\Phi,\varphi)\;=
\;2i(\frac{\omega}{c})\mid E^{(\omega)}_{0}\mid^{2}
\times\nonumber\\
&&\left( \begin{array}{c}
A_{p}F_{c}\cos\Phi\\
A_{s}\sin\Phi\\
A_{p}N^{2}F_{s}\cos\Phi
\end{array}\right)
\left( \begin{array}{cccccc}
0&0&0&\mid\;0&0&\chi^{(2)}_{xzx}\\
0&0&0&\mid\;0&\chi^{(2)}_{xzx}&0\\
\chi^{(2)}_{zxx}&\chi^{(2)}_{zxx}&\chi^{(2)}_{zzz}&\mid\;0&0&0
\end{array}\right)
\left( \begin{array}{c}
f_{c}^{2}t_{p}^{2}\cos^{2}\varphi\\
t_{s}^{2}\sin^{2}\varphi\\
f_{s}^{2}t_{p}^{2}\cos^{2}\varphi\\
2f_{s}t_{p}t_{s}\cos\varphi\sin\varphi\\
2f_{c}f_{s}t_{p}^{2}\cos^{2}\varphi\\
2f_{c}t_{p}t_{s}\cos\varphi\sin\varphi
\end{array}\right).
\end{eqnarray}
Here, $\Phi $ and $\varphi $ denote the angles of polarization
of the reflected frequency doubled and of the
incident light (see Fig. 1). $f_{c,s}$
are Fresnel coefficients and $t_{s,p}$ are the linear transmission
coefficients. The complex indices of refraction at frequencies $\omega $
and 2$\omega $ are $n\;=\;n_{1}\;+\;ik_{1}$ and $N\;=\;n_{2}\;+\;ik_{2}$.
The projections of the incident wave vector {\bf k} on the spatial coordinates
inside the medium are $f_{s}\;=\;\sin\theta/n$ and $f_{c}\;=
\;\sqrt{1-f_{s}^{2}}$. The corresponding quantities for the reflected
SHG light are $F_{s}\;=\;\sin\Theta/N$ and
$F_{c}\;=\;\sqrt{1-F_{s}^{2}}$, where $\theta $ and $\Theta $
denote the angle of incidence of the incoming light and the angle
of reflection of the reflected SHG light, respectively.
The linear transmission coefficients are given by~\cite{klaus,klausdiss,sipe}
\begin{equation}
t_{p}\;=\;\frac{2\cos\theta}{n\cos\theta+f_{c}}\;,
\;\;t_{s}\;=\;\frac{2\cos\theta}{\cos\theta + nf_{c}}
\;,\;\;T_{p}\;=\;\frac{2\cos\Theta}{N\cos\Theta+F_{c}}\;,
\;\;T_{s}\;=\;\frac{2\cos\Theta}{\cos\Theta + NF_{c}}
\;.
\end{equation}
The corresponding amplitudes $A_{p}$ and $A_{s}$ in Eq. (10) are
\[
A_{p}=\frac{2\pi T_{p}}{\cos\Theta}\;,
\]
and
\begin{equation}
A_{s}=\frac{2\pi
T_{s}}{\cos\Theta}\;,
\end{equation}
respectively.
{}From Eq. (10) one gets directly the intensity $I(\Phi,\varphi)$ of
the reflected SHG light:
$I(\Phi,\varphi)\;=\;\mid E^{(2\omega)}(\Phi,\varphi)\mid ^{2}$.
Thus, one obtains for $p$-polarized SHG with vacuum permittivity
$\varepsilon$ and vacuum permeability $\mu_{0}$
\begin{eqnarray}
I_{\alpha }(p-SH)\;&&=\;2\sqrt{\frac{\vareps_{0}}{\mu_{0}}}
\mid (2i\frac{\omega}{c})\mid^{2}\mid
E_{0}(\omega)\mid^{4}\mid A_{p}((F_{c}\chi^{(2)}_{xzx}
2f_{c}f_{s}+N^{2}F_{s}(\chi^{(2)}_{zxx}f_{c}^{2}+\chi^{(2)}_{zzz}f_{s}^{2}))
t_{p}^{2}\cos^{2}\varphi\;\nonumber \\
&&+\;
N^{2}F_{s}\chi^{(2)}_{zxx}t_{s}^{2}\sin^{2}\varphi)\mid^{2}\;,
\end{eqnarray}
and for $s$-polarized SHG
\begin{equation}
\!\!\!\!\!\!\!\!\!\!\!\!\!\!\!\!\!\!\!\!\!\!\!\!\!\!\!\!\!\!\!\!\!\!
\!\!\!\!\!\!\!\!\!\!\!\!\!\!\!\!\!\!\!\!\!\!\!\!\!\!\!\!\!\!
I_{\alpha }(s-SH)\;=\;2\sqrt{\frac{\vareps_{0}}{\mu_{0}}}
\mid (2i\frac{\omega}{c})\mid^{2}\mid
E_{0}(\omega)\mid^{4}\mid A_{s}\chi^{(2)}_{xzx}
2f_{s}t_{p}t_{s}\cos\varphi\sin\varphi\mid^{2}.
\end{equation}
Here, $I_{\alpha }(\beta-SH)$ denotes the intensity
of $\beta $-polarized reflected light
at frequency 2$\omega $ resulting from $\alpha $-polarized incoming light
($\alpha $ = $p,s,\varphi=\pi/4)$. In particular, one gets
for $p$-polarized incident and $p$-polarized reflected SHG light
\begin{equation}
I_{p}(p-SH)\;=\;2\sqrt{\frac{\vareps_{0}}{\mu_{0}}}
\mid (2\frac{\omega}{c})\mid^{2}\mid
E_{0}(\omega)\mid^{4}\mid A_{p}t_{p}^{2}(F_{c}\chi^{(2)}_{xzx}
2f_{c}f_{s}+N^{2}F_{s}(\chi^{(2)}_{zxx}f_{c}^{2}+\chi^{(2)}_{zzz}f_{s}^{2}))
\mid^{2}\;,
\end{equation}
and for $s$-polarized incident and $p$-polarized reflected SHG light
\begin{equation}
I_{s}(p-SH)\;=\;2\sqrt{\frac{\vareps_{0}}{\mu_{0}}}
\mid (2\frac{\omega}{c})\mid^{2}\mid
E_{0}(\omega)\mid^{4}\mid A_{p}N^{2}F_{s}\chi^{(2)}_{zxx}t_{s}^{2}\mid^{2}\;,
\end{equation}
and furthermore for ``mixed''-polarized (45$^{\circ }$ polarization) incident
and $s$-polarized SHG light
\begin{equation}
I_{\varphi =\pi/4}(s-SH)\;=\;2\sqrt{\frac{\vareps_{0}}{\mu_{0}}}
\mid (\frac{\omega}{c})\mid^{2}\mid
E_{0}(\omega)\mid^{4}\mid A_{s}\chi^{(2)}_{xzx}
2f_{s}t_{p}t_{s}\mid^{2}\;.
\end{equation}
The response in the latter two configurations
is determined by one single tensor element $\chi^{(2)}_{zxx}$ and
$\chi^{(2)}_{xzx}$, respectively.
Then we get for an incidence angle of 45 $^{\circ}$ for the ratio
\newline
$I_{p}(p-SH)/I_{s}(p-SH)$ the result
\begin{equation}
\frac{I_{p}(p-SH)}{I_{s}(p-SH)}\;=\;\mid\frac{(F_{c}\chi^{(2)}_{xzx}
2f_{c}f_{s}+N^{2}F_{s}\chi^{(2)}_{zzz}f_{s}^{2})^{2}t_{p}^{4}}
{N^{4}F_{s}^{2}\chi^{(2)2}_{zxx}t_{s}^{4}}\mid,
\end{equation}
where the contribution from $\chi^{(2)}_{zxx}$ has been neglected in the
numerator. Note, this tensor element has been experimentally determined
from $I_{\varphi =\pi/4}(s-SHG)$ and has been found to be at least three times
smaller than the other two tensor elements (see below). Eq. (17) holds
also for noble metals.

It follows now from Eq.(18) that $s$ and $d$ electrons contribute rather
differently to the intensity of SHG light:

First, for noble metals like Cu, Ag, Au one gets using Eq. (16) due to\\
$\chi^{(2)}_{ijl} \sim \langle s\mid i\mid s\rangle\;
\langle s\mid j\mid s\rangle\;\langle s\mid l\mid s\rangle$
and $\langle s\;\mid x\mid\;s\rangle\;=\;0$
\begin{equation}
I_{s}(p-SH)\;=\;0,
\end{equation}
if the light frequency $\omega $ is such that
$\hbar\omega \;<\;(\varepsilon_{F}\;-\;\varepsilon_{d}^{max})$
implying that no transitions involving
$d$-states occur.
Note, $\chi_{zxx}^{(2)}$ in Eq. (16) involves the matrix elements
$\langle s \mid x \mid s\rangle $ and these are nearly zero.
Hence, for noble metals
\begin{equation}
\frac{I_{s}(p-SH)}{I_{p}(p-SH)}\;\longrightarrow \;0,
\end{equation}
if the optical response does not involve $d$ state transitions.
Thus, Eqs. (18) and (20) suggest a very sensitive test of the quantity
($\varepsilon_{F}\;-\;\varepsilon_{d}^{max}$) and of many-body contributions
to this. Comparison with
results from bandstructure calculations may help to identify
correlation effects.

\noindent
If the frequency $\omega $ increases, then for
$\hbar\omega \;\geq\;(\varepsilon_{F}\;-\;\varepsilon_{d}^{max})$
also $s\longrightarrow d$ and $d\longrightarrow d$ transitions
contribute and
\begin{equation}
I_{s}(p-SH)\sim \mid\sum_{ijl}t_{s}^{2}F_{s}\chi_{zxx}^{(2)d}\mid^{2}
\;+\;\mid\sum_{ijl}t_{s}^{2}F_{s}\chi_{zxx}^{(2)s}\mid^{2}\;+\;\ldots\;.
\end{equation}
Hence, for increasing frequency $\omega $,
$I_{s}(p-SH)$ is no longer (nearly) zero, increases and
$I_{s}(p-SH)\longrightarrow I_{p}(p-SH)$ as for transition metals.

Secondly, for transition metals like Fe, Co, Ni, Cr,
we estimate in contrast to noble metals for all frequencies $\omega $, that
\begin{equation}
I_{p}(p-SH)\;\stackrel{>}{\sim}\;I_{s}(p-SH).
\end{equation}
For example, for the Ni(001) surface, using $\hbar \omega \;=\;2.056$ eV
($\lambda $ = 603 nm) and for the optical constants~\cite{jc} $n_{1}=1.98$,
$k_{1}=3.92$, $n_{2}=2.02$, and $k_{2}=2.18$, we find using Eq. (17) for
$\theta $ = 45 $^{\circ}$
\begin{equation}
0.75\leq\frac{I_{p}(p-SH)}{I_{s}(p-SH)}\leq 6.7,
\end{equation}
if we allow the relative phase of the tensor elements $\chi^{(2)}_{zzz}$ and
$\chi^{(2)}_{xzx}$ to vary between 0 and 2$\pi $.
This result gives already a reasonable account of the experimentally observed
ratio of 1.25.

Note, the estimate of the ratio $I_{p}(p-SH)/I_{s}(p-SH)$ was obtained only on
the basis of symmetry considerations for the matrix elements in
$\chi^{(2)}_{ijl}$ and by approximating these by atomic orbital matrix
elements.
Of course, for discussing systematically the nonlinear optical response
of different noble and transition metals we have to include density of states
effects resulting from the function $F_{k k^{\prime } k^{\prime
\prime }}(\omega )$. How much this matters can be seen from results for
$\chi^{(2)}_{zxx}/\chi^{(2)}_{xzx}$. This ratio should be
equal to one if $F_{k k^{\prime } k^{\prime\prime }}(\omega )$ plays no role.
This corresponds to the case of perfect Kleinman symmetry~\cite{klein}.
However, note that in $\chi^{(2)}_{ijl}$ the indices $j,l$ refer to
$\omega $ photon transitions, while $i$ belongs to the $2\omega $ photon
transition . This fact can be rather important, for example, for analyzing
differences between noble metals
and transition metals. For transition metals $\chi^{(2)}_{xjl}$ is
rather unfavorable for the SHG yield, since $s \;\longrightarrow \;s$
transitions at frequency 2$\omega $ may occur and reduce the intensity
$I_{\varphi=\pi/4}(s-SH)$ due to
$\langle s\;\mid x\mid\;s\rangle\;\ll\;\langle d\;\mid x\mid\;d\rangle $,
see for illustation Figs. 2c and 2d. In contrast, for noble metals
$\chi^{(2)}_{xjl}$ is favorable for the SHG yield,
if the transitions at frequency $\omega $ involve $s \;\longrightarrow \;s$
transitions only and  if $d \;\longrightarrow \;s$ transitions at frequency
2$\omega $ come into play. The situation is
illustrated in Figs. 2a and 2b. Thus, the deviation from Kleinman's symmetry
of the ratio $\chi^{(2)}_{zxx}/\chi^{(2)}_{xzx}$ is interesting, a
fingerprint of density of states effects, and reflects to some extent the
contribution of $d$ electrons to the optical transitions and the degree of
localization of the Fermi-level electrons. Experimentally this ratio
ranges from less than 0.013
for Al (with no $d$ electrons) over 0.085 for Cu (where
the upper $d$-band edge is 1.9 eV below the Fermi-level) and 0.33 for Au
(where the upper $d$-band edge is 1.8 eV below the Fermi-level) to 0.35 for
Ni~\cite{klaus} (with a large density of $d$- electrons at the
Fermi level). For the determination of the ratio
$\chi^{(2)}_{zxx}/\chi^{(2)}_{xzx}$ see remarks below.

As can be seen from the large range of values for the ratio
$\frac{I_{p}(p-SH)}{I_{s}(p-SH)}$ in Eq. (23), it is very important for a
quantitative analysis to include not only the
symmetry properties of the dipole matrix elements, but also
the phases, since the tensor elements $\chi^{(2)}_{ijl}$ are
complex quantities. As a result, important interference contributions may
occur and change the results obtained from a coherent superposition of the
partial intensities given by $\mid\chi^{(2)}_{ijl}\mid$ alone. For a full
analysis one has to perform an electronic calculation of $\chi^{(2)}_{ijl}$
as done previously by Pustogowa {\em et al.}~\cite{hub}. To compare theory and
experiment, the latter has to be performed for several optical geometries
to allow for a complete determination of the absolute values and phases
of $\chi^{(2)}_{ijl}$.

In determining the phases at least approximately we may proceed as follows.
To see how the two relative phases between the three
tensor elements
$\chi^{(2)}_{zzz}$, $\chi^{(2)}_{xzx}$ and $\chi^{(2)}_{zxx}$ entering
$I_{\alpha }(p-SH)$ come into play, we rewrite Eq. (13) as
\begin{eqnarray}
I_{\alpha }(p-SH)\;&&=\;\mid (2i\frac{\omega}{c})\mid^{2}\mid
E_{0}(\omega)\mid^{4}\mid A_{p}\chi^{(2)}_{zzz}([F_{c}\frac{\chi^{(2)}_{xzx}}
{\chi^{(2)}_{zzz}}
2f_{c}f_{s}+N^{2}F_{s}(\frac{\chi^{(2)}_{zxx}}{\chi^{(2)}_{zzz}}
f_{c}^{2}+f_{s}^{2})]
t_{p}^{2}\cos^{2}\varphi\;\nonumber\\
&&+\;
N^{2}F_{s}\frac{\chi^{(2)}_{zxx}}{\chi^{(2)}_{zzz}}
t_{s}^{2}\sin^{2}\varphi)\mid^{2},
\end{eqnarray}
with
\[
\frac{\chi^{(2)}_{xzx}}{\chi^{(2)}_{zzz}}\;=
\;\mid\frac{\chi^{(2)}_{xzx}}{\chi^{(2)}_{zzz}}\mid e^{i\varphi_{1}}
\]
and
\begin{equation}
\;\;\;\;\;\frac{\chi^{(2)}_{zxx}}{\chi^{(2)}_{zzz}}\;=
\;\mid\frac{\chi^{(2)}_{zxx}}{\chi^{(2)}_{zzz}}\mid e^{i\varphi_{2}}.
\end{equation}
Now, the two absolute ratios
$\mid\frac{\chi^{(2)}_{xzx}}{\chi^{(2)}_{zzz}}\mid$
and $\mid\frac{\chi^{(2)}_{zxx}}{\chi^{(2)}_{zzz}}\mid$ and the two relative
phases $\varphi_{1}$ and $\varphi_{2}$ remain to be determined.
As can be seen from Eqs. (16) and (17),
the intensities $I_{\varphi =\pi/4}(s-SHG)$ and $I_{s}(p-SHG)$
depend exclusively on one single nonlinear tensor element each,
$\chi^{(2)}_{xzx}$ and $\chi^{(2)}_{zxx}$, respectively.
Thus, the intensity ratio
\[
\frac{I_{\varphi =\pi/4}(s-SH)}{I_{s}(p-SH)}\;,
\]
which may be taken from experiment, gives directly the absolute ratio
$\mid\chi_{xzx}^{(2)}/\chi_{zxx}^{(2)}\mid $.
Using for example for $I_{\varphi =\pi/4}(s-SH)/I_{s}(p-SH)$ the experimental
value and the appropriate linear optical constants of Ni one gets
\begin{equation}
\mid\frac{\chi_{xzx}^{(2)}}{\chi_{zxx}^{(2)}}\mid \;=\;1/0.35\;\approx 2.8.
\end{equation}
Similarly, we calculate (1/0.013) for Al, (1/0.085) for Cu, and (1/0.33) for
Au.

To obtain the remaining relative phases $\varphi_{1}$ and $\varphi_{2}$, and
the absolute ratio $\mid\chi_{xzx}^{(2)}/\chi_{zzz}^{(2)}\mid $ one may
use the results for the $p$-polarized SHG-yield for incident $p$ polarized
light. Note, according to Eq. (15),
$I_{p}(p-SH)$ involves a superposition of all three independent
nonvanishing tensor elements $\chi_{zzz}^{(2)}$, $\chi_{xzx}^{(2)}$ and
$\chi_{zxx}^{(2)}$.
Keeping $F_{k k^{\prime} k^{\prime \prime}}(\omega ,2\omega )$ fixed, we may
calculate all surface transition matrix elements entering $\chi_{ijl}^{(2)}$
within an basis set of atomic $4s$, $4p$, and $3d$ wave functions.
Using these matrix elements (see table I), the decomposition of the
nonlinear tensor $\chi ^{(2)}_{ijl} $ in Eq. (6) and Eq. (8) together with
the partial $p$ polarized SHG yields for transition metals in table II,
gives us for Ni the absolute ratio
\begin{equation}
\mid\frac{\chi_{xzx}^{(2)}}{\chi_{zzz}^{(2)}}\mid\;=\;\frac{3}{5}.
\end{equation}
It follows then from Eqs. (16) and (17) that the remaining absolute ratio
is
\begin{equation}
\mid\frac{\chi_{zxx}^{(2)}}{\chi_{zzz}^{(2)}}\mid\;=\;
\mid\frac{\chi_{zxx}^{(2)}}{\chi_{xzx}^{(2)}}\mid
\mid\frac{\chi_{xzx}^{(2)}}{\chi_{zzz}^{(2)}}\mid\;=\;0.35\;\cdot\;\frac{3}{5}
\;=\;0.21.
\end{equation}
Using then Eqs. (24), (27) and (28) and the experimental results
for $I_{\alpha }(p-SH)$ one can determine the
remaining two relative {\em phases} of the three tensor elements.
We find good agreement with experiment, if we use (for Ni)
\[
\chi_{xzx}^{(2)}\;=\;\chi_{zzz}^{(2)}e^{i1.95\pi}\;,
\]
and
\begin{equation}
\chi_{zxx}^{(2)}\;=\;\chi_{zzz}^{(2)}e^{i0.505\pi}\;,
\end{equation}
see results presented in Fig. 3.
Thus, we have completely determined the absolute ratios and the relative phases
of all non-vanishing elements of the nonlinear tensor $\chi^{(2)}_{ijl}$
for surfaces of fcc crystals.

It is important to note, that this procedure shows how the polarization
dependence of the SHG yield can be used to determine the various tensor
elements of the nonlinear susceptibility $\chi_{ijl}^{(2)}$. Furthermore, it
shows also how in turn the microscopic susceptibility $\chi_{ijl}^{(2)}$
determines the polarization dependence of the SHG yield.

Another interesting fingerprint of the electronic structure is the behavior of
$I_{\varphi}(s-SH)$.
While it follows straightforwardly from symmetry and matrix elements, see Eq.
(14), that
\begin{equation}
I_{p}(s-SH)\;\approx\;I_{s}(s-SH)\;\approx\;0\;,
\end{equation}
the behavior of the ratio $I_{\varphi=45^{\circ}}(s-SH)/I_{s}(p-SH)$, in
particular the observation~\cite{klaus,klausdiss}
\begin{equation}
(\frac{I_{\varphi=45^{\circ}}(s-SH)}{I_{s}(p-SH)})_{Cu}\;>\;
(\frac{I_{\varphi=45^{\circ}}(s-SH)}{I_{s}(p-SH)})_{Au}
\end{equation}
requires a more careful theoretical analysis, including in the discussion
of $\chi^{(2)}_{ijl}$ not only the dipole matrix elements, but again also the
energy denominators. Note, $\chi ^{(2)}_{ijl}$ can be written as
\begin{equation}
\chi _{ijl}^{(2)}\;=\;const.
\sum_{{\bf k},{\bf k^{\prime}},{\bf k^{\prime \prime}}}
\langle k\mid i\mid k^{\prime }\rangle\langle k^{\prime }\mid j\mid
k^{\prime \prime }\rangle
\langle k^{\prime \prime }\mid l\mid k\rangle
\frac{\chi^{(1)}_{{\bf k^{\prime \prime }k^{\prime}}}
-\chi^{(1)}_{{\bf k^{\prime }k}}}
{E_{\bf k^{\prime \prime}}-E_{\bf k}-2\hbar \omega}\;,
\end{equation}
where $\chi ^{(1)}$ denotes essentially the linear susceptibility,
but without matrix elements~\cite{hub}.
Thus,
\[
F_{{\bf k},{\bf k^{\prime}},{\bf k^{\prime
\prime}}}(\omega , 2\omega )\;
=\;const.\frac{\chi^{(1)}_{{\bf k^{\prime \prime }k^{\prime}}}
-\chi^{(1)}_{{\bf k^{\prime }k}}}
{E_{\bf k^{\prime \prime}}-E_{\bf k}-2\hbar \omega}\;.
\]
Hence, $I_{\alpha }(s -SH)$ reveals in addition to the matrix elements
the behavior of the function $F_{k k^{\prime } k^{\prime
\prime }}(\omega ,2\omega )$ involving the density of states. Note, the
$\omega $ dependence must follow from  $F_{k k^{\prime } k^{\prime
\prime }}(\omega ,2\omega )$ and will be more general than the $\omega $
dependence deduced from the jellium.\\

To understand now the result given in Eq. (31), note that
$(I_{\varphi=45^{\circ}}(s-SH))_{Cu}\;>\;(I_{\varphi=45^{\circ}}(s-SH))_{Au}$,
since the density of $p$ states above $\varepsilon _{F}$ is larger for Cu,
while $(I_{s}(p-SH))_{Cu}\;<\;(I_{s}(p-SH))_{Au}$, since the density of $d$
states at $\hbar \omega $ below $\varepsilon _{F}$ is larger for
Au~\cite{papa}.

Note, the ratio
\[
\frac{I_{\varphi=45^{\circ}}(s-SH)}{I_{s}(p-SH)}
\]
probes essentially the density of $d$ electrons at an energy of $\hbar
\omega $ below the Fermi-level and of $p$ states above $\varepsilon _{F}$, see
for illustration Fig. 2.
As a consequence, it follows from Eqs. (16) and (17) that the ratio
$\chi^{(2)}_{zxx}/\chi^{(2)}_{xzx}$
which describes the possibility of $d$ electron excitations by optical
transitions
increases from the jellium type metal Al via the noble metals Au and Cu to
the transition metal Ni~\cite{klaus,klausdiss}.

Thus, it is interesting to note that the combination of the
symmetry properties of the (real) transition matrix elements and the density
of states effects will account for a quantitative understanding of the
SHG yield. Furthermore, since the tensor elements $\chi ^{(2)}_{ijl}$ are
complex quantities, their phases cause important interference
contributions to the coherent superposition of the partial SHG yields
resulting from the single tensor elements alone. An {\em electronic}
theory along the lines of Pustogowa {\em et al.}~\cite{hub} is required
for the calculation of the full complex SHG tensor $\chi ^{(2)}_{ijl}$
to explain the experiments quantitatively and in detail. It is clear from
Eqs. (10) -- (17) that experiments have to be carried out for several optical
geometries to sort out for comparison with theory the tensor elements and to
allow for a complete determination of their {\em absolute values} and
{\em phases}.

The physics described by our theory is evident also from the results shown in
table I. Here, we have estimated the dipole matrix elements which determine
largely $\chi^{(2)}_{ijl}$ by using atomic $s(p)$ and $d$-electron
wavefunctions. Decomposing $\chi^{(2)}_{ijl}$ according to Eq. (6), it follows
\begin{equation}
\chi^{(2,s)}_{zzz}\;\gg\;\chi^{(2,s)}_{zxx},
\end{equation}
since $\langle s\;\mid x\mid\;s\rangle\;\approx\;0$, and
\begin{equation}
\chi^{(2,d)}_{zzz}\;\stackrel{>}{\sim}\;\chi^{(2,d)}_{zxx},
\end{equation}
since
$\langle d\;\mid z\mid\;d\rangle\;\stackrel{>}{\sim}
\;\langle d\;\mid x\mid\;d\rangle $.
Generally, for transition metals, $\chi^{(2,d)}_{ijl}$ is largest by far.
For noble metals, $\chi^{(2,d)}_{ijl}$ begins to dominate over
$\chi^{(2,s)}_{ijl}$ as $\hbar \omega $ increases and becomes larger than
the energy distance of the $d$ states from the Fermi energy $\varepsilon_{F}$,
$\hbar\omega \;>\;(\varepsilon_{F}\;-\;\varepsilon_{d}^{max})$.
Furthermore, approximately $I_{p}(p-SH)$ is essentially given by
$\chi^{(2)}_{zzz}$ and $\chi^{(2)}_{xzx}$, whereas $I_{s}(p-SH)$
is determined by $\chi^{(2)}_{xzx}$ alone. In the case of $s$-polarized
SH light $I_{s}(s-SH) \; = \;I_{s}(p-SH)\; = \;0$. And
$I_{\varphi =\pi/4}(s-SH)$ is given by $\chi^{(2)}_{xzx}$ alone.
Hence, using table I, we find quite generally that:\\
(a) for {\em noble metals} such as Cu, Au, Ag\\
\indent\indent $I_{p}(p-SH)\;\gg\;I_{s}(p-SH)\;,\;\;$
if $\hbar\omega \;<\;(\varepsilon_{F}\;-\;\varepsilon_{d}^{max})$\\
\indent\indent $I_{p}(p-SH)\;\approx
\;\mid\chi^{(2,s)}_{ijl}\mid^{2}\;+\;\mid\chi^{(2,int)}_{ijl}\mid^{2}\;,\;\;$
if $\hbar\omega \;>\;(\varepsilon_{F}\;-\;\varepsilon_{d}^{max})$.\\
\indent\indent $I_{p}(p-SH)\;\longrightarrow\;I^{trans. met.}_{p}(p-SH)\;,\;\;
\hbar \omega $ large.\\
\\
(b) for {\em transition metals} such as Ni, Co, Fe, Cr\\
\indent\indent $I(p-SH)\;\approx \;\mid\chi^{(2,d)}_{ijl}\mid^{2}$,\\
and\\
\indent\indent $I_{p}(p-SH)\;\stackrel{>}{\sim} \;2I_{s}(p-SH)$.\\
This completes then the theoretical analysis of the polarization dependence
of the SHG yield and shows how it reveals the electronic structure at surfaces
and interfaces.
\\
\\
{\bf III. Results and Discussion}

To demonstrate the different polarization dependence of the SHG yield for
noble metals and transition metals, we show in Fig. 3 numerical results for
the polarization dependence of the $p$ polarized SHG yield $I_{\varphi}(p-SH)$.
Using Eq. (24), we calculate $I_{\varphi}(p-SH)$ for Ni and Cu where the
$p$-polarized SHG-yield at $p$-polarized incident polarization has been
normalized to unity. The
angle $\varphi $ characterizes the polarization of the incident light. We show
results for Ni and Cu
at angle of incidence $\theta \;=\;45\;^{\circ}$ corresponding to the
experimental situation~\cite{klaus,klausdiss}. The absolute ratios and
relative phases of the tensor elements are listed in table III as well as the
used linear optical constants at frequencies $\omega $ and 2$\omega $.

First, we refer to results for Ni given by the topmost curve. These are typical
for {\em transition metals}.
For Ni we used an incident light wavelength of 603 nm. The results are
in perfect agreement with experiment~\cite{klaus,klausdiss}.
Despite the fact that we have fitted the phases $\varphi_{1}$ and $\varphi_{2}$
to experiment this is remarkable, since the ratio
$\mid\chi_{xzx}^{(2)}/\chi_{zzz}^{(2)}\mid$ has been obtained using only
the matrix elements calculated with atomic wave functions.
As mentioned the line shape is characteristical for
transition metals. Similar results are expected for Cr and for Fe and
Co. The SHG yield for $s$ and $p$ polarized incident light is nearly the same
in transition metals due to the {\em isotropy} of the transition matrix
elements for $d$ electrons
\[
\langle d\mid z\mid d\rangle\;\approx\;\langle d\mid x\mid d\rangle.
\]
Slight intensity minima occur at polarization angles of 45$^{\circ}$ and
135 $^{\circ}$. These arise from the interference of the tensor elements
$\chi _{ijl}^{(2)}$ and the Fresnel factors, see Eq. (13). The intensity
profile
is always symmetric around $\varphi \;=\;90^{\circ }$ and for transition metals
approximately symmetric around $\varphi\;=\;45^{\circ}$ and
$\varphi\;=\;135^{\circ}$.

Secondly, to demonstrate the typical frequency dependence of the SHG yield as a
function of incident polarization for {\em noble metals} (N. M.), results for
Cu are given in Fig. 3 by the curves (a), (b), (c) . Note, the incident light
polarization dependence of the SHG yield is found to be very different for
noble metals such as Cu than for transition metals (T. M.). The
$\varphi $ = 90 $^{\circ}$ symmetry survives. The $p$-polarized SHG yield falls
off to one fourth or less at incident $s$-polarization compared to its
(normalized) value at incident $p$ polarization, due to the strong
{\em anisotropy} of the transition matrix elements for $s$ electrons
\[
\langle s\mid z\mid s\rangle\;\gg\;\langle s\mid x\mid s\rangle\;\approx\;0.
\]
The Cu curves (a) and (c) refer to wavelenghts
exciting and non-exciting the Cu $d$ electrons, respectively. The results are
in very good agreement with experimental ones for the frequency dependent
SHG yield obtained for Cu by Petrocelli {\em et al.}~\cite{petro}.
While the curve (a), which is for an incident light wavelength of 516.6 nm,
still shows shallow minima at 63 $^{\circ}$ and 117 $^{\circ}$,
the curve (c), which is for an incident light wavelength of 652.5 nm,
falls off monotonously to zero at incident $s$-polarization (90 $^{\circ}$).
These minima are due to the phases of the complex nonlinear tensor elements
giving rise to interference contributions as well as due to the complex optical
constants. These interferences will be absent if the $d$ band of Cu can
no longer be excited for larger wavelenghts. Note, the shape of curve (a)
would also be characteristic of Au at 603 nm as is observed in the
experiment by B\"ohmer {\em et al.}~\cite{klaus,klausdiss}. These results
indicate that
\[
(I_{s}(p-SH))_{N.M.}\;\stackrel{\omega }{\longrightarrow }
\;(I_{s}(p-SH))_{T.M.}
\]
for increasing frequency $\omega $ as claimed previously on theoretical
grounds.

Note, that experimentally
$I_{s}(p-SH)$ is larger for Au than for Cu. The
experiments~\cite{klaus,klausdiss} are performed for $\hbar\omega\;=\;2.056$
eV.
Then, $\hbar\omega
\;>\;(\varepsilon_{F}\;-\;\varepsilon_{d}^{max})$,
according to bandstructure calculations by Papaconstantopoulos~\cite{papa}
yielding $(\varepsilon_{F}\;-\;\varepsilon_{d}^{max})\;\approx\; 1.8$ eV for
Au and 1.9 eV for Cu.
Hence, one would
expect for both metals Cu and Au already larger values for $I_{s}(p-SH)$.
However, note
that due to many-body band narrowing
$(\varepsilon_{F}\;-\;\varepsilon_{d}^{max})$ is for both metals
somewhat larger than what
results from bandstructure calculations. However, since
$(\varepsilon_{F}\;-\;\varepsilon_{d}^{max})_{Au}
\;<\;(\varepsilon_{F}\;-\;\varepsilon_{d}^{max})_{Cu}$ one also expects
in agreement with experiment
\[(I_{s}(p-SH))_{Au}\;>\;(I_{s}(p-SH))_{Cu}\]
at energy $\hbar \omega \;=\;2.056 eV$.

The curve (b) corresponds also to Cu at a wavelength of 652.5 nm as
curve (c). For this curve (b), however, only the optical constants have been
adapted to this wavelength, whereas the nonlinear tensor elements of wavelength
516.6 nm were kept as in curve (a).
At $\varphi\;=\;90^{\circ }$, this curve reaches the same height as the Cu
curve (a), but it shows deep minima at $\varphi\;=\;54^{\circ }$ and at
$\varphi \;=\;126^{\circ }$. Thus, this curve (b) clearly demonstrates that
for the
correct theoretical description of the frequency and polarization dependence
of the SHG yield the frequency dependence
of the nonlinear tensor elements is as essential as the frequency dependence
of the optical constants at frequencies $\omega $ and 2$\omega $.

In summary, the polarization dependence of the SHG-yield
shows characteristic curve shapes which allow for a clear distinction
between the electronic structure of noble metal and transition metal surfaces.
As claimed, SHG is simply a very sensitive probe of the electronic structure.

One very interesting application of our theory for the frequency dependence of
the SHG-yield is to transition metal oxides~\cite{note2}.
Note, late 3$d$ transition metal oxides, in particular NiO, are charge transfer
insulators. NiO has a gap of 4 eV separating the
filled oxygen $2p$ band from the empty upper Hubbard portion of the Ni 3$d$
band, see Fig. 4 for illustration.
These transition metal oxides are of particular interest
due to the strong electron correlations.
Thus, the frequency dependent SHG could give some important information about
the energy gap of the transition metal oxides and the width of the upper
Hubbard band and thus about correlations. Fig. 4 illustrates the electronic
structure of NiO, for example.

Accordingly, one expects for example for NiO at $\hbar \omega \;<\;\Delta $,
where $\Delta $ refers to the charge transfer gap, no SH yield. As a
consequence, oxidization of the Ni surface (in air) would not essentially
change the experimental results for Ni, since the Ni below the oxidized surface
would mainly contribute to the SHG yield. Similarly, this conclusion can also
be
drawn for Cu. Actually, for Cu, the irrelevance of the oxide layer
was shown by Petrocelli {\em et al.}~\cite{petro} from the agreement of the
experimental data in air and in ultrahigh vacuum.
Furthermore, since the SHG-yield would disappear if
$\hbar \omega \;\stackrel{>}{\sim} \;\Delta $ and $\hbar \omega \;\geq \;W $,
nonlinear optics could be used to determine the width $W$ of the upper Hubbard
band.

We have neglected hybridization between the oxygen 2$p$ states and the
majority 3$d$ states of Ni. However note that at $\varepsilon_{F}$ and above
only minority $d$ states are present.

In view of the different {\em magnetic} properties of the NiO-vacuum interface
and of the Ni-NiO interface, the nonlinear magneto-optical Kerr-effect
(NOLIMOKE) may provide essential additional information regarding the
electronic
structure and the location of the electronic contribution to the SHG signal.
Very interestingly, the nonlinear magneto-optical Kerr
effect should vary characteristically with light frequency $\omega $,
since the $d$-electron spin polarization varies throughout the
$d$-electron bands of Ni, whereas no nonlinear Kerr-signal is expected
for antiferromagnetic NiO. Thus, in the range of optical frequencies, only
minority $d$-electrons of Ni at the Ni-NiO interface contribute to the NOLIMOKE
signal.

Finally, it is of interest to apply our theory also to surface and
interface states having generally different $s,p$ and $d$
characters  and spin-polarization.

We have neglected contributions to the SHG yield of higher than electric dipole
order. However, we estimate them to be smaller by a factor of $\overline{\alpha
 }^{2} \approx (1/137)^{2} = 5 \cdot 10^{-5}$ for the magnetic dipole and of
(2$\pi a_{0}/\lambda )^{2} \approx 2.5\cdot 10^{-7}$ for the electric
quadrupole contribution. Here, $\overline{\alpha }$ denotes the fine structure
constant, and $a_{0}$ is Bohr's radius. The resulting smallness cannot be
compensated by the skin depth factor. The implied interface sensitivity of SHG
is supported by the agreement of our theory with experiment.

The $\omega $ dependence of the SHG yield follows from the function
$F_{k k^{\prime } k^{\prime \prime }}(\omega ,2\omega )$, see Eq. (32).
It could be interesting to sort out this in limiting cases in order to compare
with results from continuum electrodynamics and jellium
approximation~\cite{petro}.

In summary, our theory shows that the nonlinear optical response at transition
and noble metal surfaces and interfaces is a sensitive fingerprint of the
electronic structure. Both, matrix elements and the density of states affect
characteristically the intensity of the SHG 2$\omega $ reflected light. While
a more detailed numerical analysis is desirable, it will not change our
qualitative arguments based on general physical facts.

\acknowledgements
The authors gratefully acknowledge many stimulating discussions with
Professor E. Matthias.

\newpage
\begin{center}
\begin{table}
\begin{minipage}{15cm}
\caption{Non-dipole transition matrix elements at the surface
involved in the calculation of $\chi ^{(2)}_{ijl}$ and $I(SH)$,
respectively.}
\end{minipage}
\begin{minipage}{9cm}
\begin{tabular}{|l|l|}\hline
$\;\;\;\;\;\;\;$matrix element $\;\;\;\;\;\;\;\;\;\;\;\;\;\;\;\;\;\;\;$
&value \\ \hline \hline
$\langle 4s\mid x\pm iy\mid 4s\rangle $ & $0$ \\ \hline
$\langle 4s\mid z\mid 4s\rangle $ & $-\frac{a_{B}}{Z}$\\ \hline
$\langle 4p_{m=\pm 1}\mid z\mid 4p_{m=\pm 1}\rangle $ &
$-\frac{69}{16}\frac{a_{B}}{Z}$ \\ \hline
$\langle 4p_{m=0}\mid z\mid 4p_{m=0}\rangle $ &
$-\frac{69}{8}\frac{a_{B}}{Z}$ \\ \hline
$\langle 3d_{m=\pm 2}\mid z\mid 3d_{m=\pm 2}\rangle $ &
$-\frac{105}{64}\frac{a_{B}}{Z}$ \\ \hline
$\langle 3d_{m=\pm 1}\mid z\mid 3d_{m=\pm 1}\rangle $ &
$-\frac{105}{32}\frac{a_{B}}{Z}$ \\ \hline
$\langle 3d_{m=0}\mid z\mid 3d_{m=0}\rangle $ &
$-\frac{105}{32}\frac{a_{B}}{Z} $ \\ \hline
$\langle 3d_{m=0}\mid z\mid 4s\rangle $ &
$0.7005\frac{\sqrt{5}}{16}\frac{a_{B}}{Z}$ \\ \hline
$\langle 4p_{m=\pm 1}\mid x\pm iy\mid 4p_{m=0} \rangle $ &
$-\frac{69}{8\sqrt{2}}\frac{a_{B}}{Z} $ \\ \hline
$\langle 4p_{m=0}\mid x\pm iy\mid 4p_{m=\mp 1}\rangle $ &
$-\frac{69}{8\sqrt{2}}\frac{a_{B}}{Z}$ \\ \hline
$\langle 3d_{m=\pm 2}\mid x\pm iy\mid 3d_{m=\pm 1}\rangle $ &
$-\frac{105}{32}\frac{a_{B}}{Z }$  \\ \hline
$\langle 3d_{m=\mp 1}\mid x\pm iy\mid 3d_{m=\mp 2}\rangle $ &
$-\frac{105}{32}\frac{a_{B}}{Z} $ \\ \hline
$\langle 3d_{m=\pm 2}\mid x\pm iy\mid 4s\rangle $ &
$\frac{0.7005}{8}\sqrt{\frac{15}{2}}\frac{a_{B}}{Z} $ \\ \hline
$\langle 3d_{m=\pm 1}\mid x\pm iy\mid 3d_{m=0}\rangle $ &
$0$ \\ \hline
$\langle 3d_{m=0}\mid x\pm iy\mid 3d_{m=\mp 1}\rangle $ &
$0$ \\ \hline
\end{tabular}
\end{minipage}
\label{table1}
\end{table}
\begin{table}
\begin{minipage}{15cm}
\caption{Values for the $p$-polarized partial SHG yields
$I^{(s,d,int)}_{s,p}(p-SH)$ of $d$ and $s$
electrons and interference contributions.}
\end{minipage}
\begin{minipage}{9cm}
\begin{tabular}{|l|l|}\hline
$\;\;\;\;\;\;\;$ partial SHG yield $\;\;\;\;\;\;\;\;\;\;\;\;\;\;\;\;\;\;\;$
& value \\ \hline \hline
$I^{(2)(d)}_{p-in}(p-SHG) $ & $25(\frac{105}{32})^{6}(\frac{a_{B}}{Z})^{6}$
\\ \hline
$I^{(2)(d)}_{s-in}(p-SHG) $ & $ 9(\frac{105}{32})^{6}(\frac{a_{B}}{Z})^{6}$
\\ \hline
$I^{(2)(s)}_{p-in}(p-SHG) $ & $(\frac{a_{B}}{Z})^{6} $
\\ \hline
$I^{(2)(s)}_{s-in}(p-SHG) $ & $0$
\\ \hline
$I^{(2)(int)}_{p-in}(p-SHG) $ & $0.015152(\frac{a_{B}}{Z})^{6}$
\\ \hline
$I^{(2)(int)}_{s-in}(p-SHG) $ & $0.052908(\frac{a_{B}}{Z})^{6}$
\\ \hline
\end{tabular}
\end{minipage}
\label{table2}
\end{table}
\end{center}
\begin{minipage}{15cm}
\begin{table}
\caption{Values for the absolute values and relative phases of the nonlinear
tensor elements $\chi ^{(2)}_{ijl}$ and for the optical constants $n_{i}$,
$k_{i}$ of Ni and Cu corresponding to the four curves of Fig. 3 ($\lambda $
denotes the wavelength of the incident light). The optical constants for Ni are
taken from ref. [10] and for Cu from ref. [14].}
\begin{tabular}{|l|l|l|l|l|}\hline
& Ni ($\lambda $=603 nm)& Cu ($\lambda $=516.6 nm) & Cu ($\lambda $=652.5 nm)
& Cu ($\lambda $=652.5 nm) \\
& & curve (a) & curve (b) & curve (c) \\ \hline \hline
$\mid\frac{\chi^{(2)}_{xzx}}{\chi^{(2)}_{zzz}}\mid$ & 0.60 & 0.363& 0.363 &
0.0363
\\ \hline
$\mid\frac{\chi^{(2)}_{zxx}}{\chi^{(2)}_{xzx}}\mid$ & 0.35 & 0.085 & 0.085&
0.085
\\ \hline
phase $\varphi_{1}$ & 1.945 $\pi $ & 1.02 $\pi $ & 1.02 $\pi $ & 1.02 $\pi $
\\ \hline
phase $\varphi_{2}$ & 0.505 $\pi $ & 1.34 $\pi $ & 1.34 $\pi $ & 1.34 $\pi $
\\ \hline
$n_{1}(\omega )$ & 1.98 & 1.12 & 0.214 & 0.214
\\ \hline
$k_{1}(\omega )$ & 3.92 & 2.60 & 3.67 & 3.67
\\ \hline
$n_{2}(2\omega )$ & 2.02 & 1.53 & 1.34 & 1.34
\\ \hline
$k_{2}(2\omega )$ & 2.18 & 1.71 & 1.81 & 1.81
\\ \hline
\end{tabular}
\label{table3}
\end{table}
\end{minipage}
\newpage
\noindent
\begin{figure}
\caption{The polarization and geometry of the incoming $\omega $ and reflected
2$\omega $ light, respectively. $\varphi $ and $\Phi $ are the polarizations
of the incident light and the reflected frequency-doubled photons.
$\varphi =0^{\circ }$ corresponds to $p$ polarization and
$\varphi =90^{\circ }$ to $s$ polarization. $\theta $ denotes the angle of
incidence. The crystal axes
$x$ and $y$ are in the crystal-surface plane whereas $z$ is parallel to the
surface normal.}
\label{fig1}
\end{figure}
\begin{figure}
\caption{Illustration of second harmonic generation from noble metals
[(a) and (b)] and transition metals [(c) and (d)]. For noble metals,
in case (a) no $d$ electrons can be optically excited, which in contrast is
possible in case (b). In case (c) for transition metals, predominantly $d$
electrons contribute to the SHG yield, whereas in case (d) the excitation
starts from the $s$ band.}
\label{fig2}
\end{figure}
\begin{figure}
\caption{Numerical results for the SHG yield $I_{\varphi}(p-SH)$ for Ni and
for Cu, curves (a), (b), and (c), using Eq. (24).
$\varphi $ refers to the polarization of the incident light. For input
parameters see text and table III.}
\label{fig3}
\end{figure}
\begin{figure}
\caption{Scheme of the electronic structure of NiO and Ni. $W$ is the width of
the upper Hubbard band, $\Delta $ is the charge transfer gap, $U$ denotes the
on-site Coulomb repulsion, $e_{g}$ and $t_{g}$ are the two and
threedimensional represenstations of $d$ electron states in cubic environment.}
\label{fig4}
\end{figure}

\begin{references}
\bibitem[1] {bloem} N. Bloembergen, R. K. Chang, S. S. Jha, and C. H. Lee,
Phys. Rev. {\bf 174}, 813 (1968); P. Guyot-Sionnest, W. Chen, and
Y. R. Shen, Phys. Rev. B {\bf 33}, 254 (1986).
\bibitem[2] {liebsch} A. Liebsch and W. L. Schaich, Phys. Rev. B {\bf 40},
5401 (1989).
\bibitem[3] {hub} W. H\"ubner, Phys. Rev. B {\bf 42}, 11553 (1990);
U. Pustogowa,
W. H\"ubner, and K. H. Bennemann, Phys. Rev. B {\bf 48}, 8607 (1993).
\bibitem[4] {petro} G. Petrocelli, S. Martellucci, and R. Francini,
Appl. Phys. A {\bf 56}, 263 (1993).
\bibitem[5] {klaus} K. B\"ohmer, E. Matthias, W. H\"ubner, and K. H.
Bennemann, to be published (1994).
\bibitem[6] {klausdiss} K. B\"ohmer, thesis (FU Berlin, 1994).
\bibitem[7] {note} At the surface and at interfaces also transitions with
$\Delta m =0,\pm 1$ and $\Delta l =0,\pm 2$ are allowed since
the perpendicular inversion symmetry is broken.
\bibitem[8] {note1}
$\chi _{ijl}^{(2)}\;=\;const
\sum_{{\bf k},{\bf k^{\prime }},{\bf k^{\prime \prime }}}
\langle k\mid i\mid k^{\prime }\rangle\langle  k^{\prime }\mid j\mid
 k^{\prime \prime }\rangle
\langle k^{\prime \prime }\mid l\mid k\rangle
\frac{\chi^{(1)}_{{\bf k^{\prime \prime }k^{\prime}}}
-\chi^{(1)}_{{\bf k^{\prime }k}}}
{E_{\bf k^{\prime \prime}}-E_{\bf k}-2\hbar \omega}$
see also below and details in ref. 3.
\bibitem[9] {sipe} J. E. Sipe, D. J. Moss, and H. M. van Driel,
Phys. Rev. B {\bf 35}, 1129 (1987).
\bibitem[10] {jc} P. B. Johnson and R. W. Christy,
Phys. Rev. B {\bf 9}, 5056 (1974).
\bibitem[11] {klein} D. A. Kleinman, Phys. Rev. {\bf 126}, 1977 (1962).
\bibitem[12] {papa} D. A. Papaconstantopoulos,
{\em Handbook of the Bandstructure of Elemental Solids} (Plenum 1986).
\bibitem[13] {note2} Ni forms in air a thin antiferromagnetic and
passivating oxide on its surface, probably two monolayers in thickness.
\bibitem[14] {hand} E. D. Palik (Ed.), {\em Handbook of Optical Constants
in Solids}, Academic, New York (1985).
\end{references}
\end{document}